\documentclass[aps,prd,preprint,draft]{revtex4}
\begin{document} 
\title{ \bf Gravitational Waves on Conductors}
\author{A. Lewis Licht}
\affiliation{Dept. of Physics\\University of Illinois at 
Chicago\\Chicago, Illinois 60607\\licht@uic.edu}

\begin{abstract}
We consider a gravitational wave impinging on 
either a normal or a super-conductor.  Fermi normal coordinates are 
derived for waves of arbitrary frequency.  The transverse-traceless 
wave amplitude expressed in 
these coordinates is used to derive the gravitationally induced 
electromagnetic fields in the conductor. In the superconducting case, 
using a time-dependent Ginzburg-Landau effective Lagrangian, we derive 
the perturbations in supercarrier density and phase.  As examples, we 
calculate the outward propagating EM 
waves produced by a low frequency gravitational wave impinging on a small 
sphere and by a high frequency wave normally incident on a large disk. 
We estimate for both targets the GW to EM conversion efficiencies and also 
the magnitude of the superconductor's phase perturbation. 
\end{abstract}
\maketitle
\section{Introduction}
There has been interest expressed lately in the possible use of a 
superconductor as a transducer between high frequency gravitational 
and electromagnetic waves\cite{chaio}. In the following we 
investigate the effect of a gravitational wave of arbitrary frequency 
on both normal and super-conductors. We attempt to make a minimum of 
special assumptions, however we do use time dependent 
Ginzburg-Landau equations for the superconducting case, which may 
only be strictly valid for temperatures near the superconducting 
transition. The superconductor is a kind of ``generic'' superconductor, 
with
Ginzburg-Landau length taken as equal to the London penetration 
depth, about 100 nm. We also assume there are no normal 
electrons present in the superconductor, a highly idealised 
condition.  Only the direct effect of the 
gravitational wave on the conducting particles is considered.  The possibility of 
electrostrictive effects are ignored, but could conceivably be 
important\cite{z}.

In section II we find the Fermi normal coordinate system for an 
object in free-fall in a gravitational wave.  The wavelength could be 
short compared to the object's size.  In section III the single 
particle Hamiltonian for a non-relativistic particle in the presence 
of gravitational and electromagnetic waves is derived and used to 
obtain general expressions for the gravitationally induced fields in both 
normal and super-conductors. As an example, in section IV 
the electromagnetic wave generated by a plane, long 
wavelength gravitational wave impinging on a sphere is derived.  
We show that in this case the outgoing EM wave is the same for both 
normal and super-conductor, and the energy conversion efficiency is 
extremely small.  We also show that for typical estimates of the 
gravitational wave intensity, the phase perturbation in the 
case of the superconductor is very sensitive to the presence of the 
gravitational wave and might be measurable.  In section V we consider 
a short wavelength plane gravitational wave normally incident on a 
large, thin disk.  We show that in this case the energy conversion 
efficiency is still extremely small. For reasonable estimates of the 
gravitational wave amplitude the magnitude of the phase perturbation is 
unchanged.  

\section{The Gravitational Wave}

\subsection{Transverse-Traceless Gauge}

We start with the coordinate metric for a gravitational wave in 
transverse-traceless gauge\cite{weinberg},
\begin{equation}
\tilde{g}_{\mu \nu} = \eta_{\mu \nu} + \tilde{h}_{\mu \nu}  \; ,
\end{equation}
where
\begin{equation}
\partial_{\mu}(\tilde{h}^{\mu}_{\nu} - 
\frac{\delta^{\mu}_{\nu}}{2}\tilde{h}^{\alpha}_{\alpha})  \;  ,
\end{equation}
\begin{equation}
\partial^{\alpha}\partial_{\alpha}\tilde{h}_{\mu \nu} = 0   \;  ,
\end{equation}
and has the plane wave solutions:
\begin{equation}
\tilde{h}_{\mu \nu} = R[\varepsilon_{\mu \nu}e^{ikx}]   \;  ,
\end{equation}
where $\varepsilon_{\mu \nu}$ is a constant tensor.  We will take the 
wave as propagating along the z-axis, and then, in complex notation, 
the wave can be expressed as 
\begin{equation}
\tilde{h}_{i j} = [h_{+ij} + h_{\times ij}]e^{-ik(t - z)}   \;  ,
\end{equation}
where
\begin{equation}
h_{+11} = - h_{+22} = h_{+}    \;  ,      h_{\times 12} = h_{\times 21} 
= h_{\times}   \;  ,
\end{equation}
all other components being zero.

The curvature tensor is, in first order,
\begin{equation}
R_{\mu \lambda \kappa \nu} = \frac{1}{2}[\tilde{h}_{\lambda \nu , \mu 
\kappa} - \tilde{h}_{\mu \nu , \lambda \kappa} - \tilde{h}_{\lambda 
\kappa , \mu \nu} + \tilde{h}_{\mu \kappa , \lambda\nu}]  \; .
\end{equation}
Which gives, in TT gauge,
\begin{equation}
R_{0101} = R_{3131} = - R_{0202} = - R_{3232} = - R_{0131} = R_{0232} 
= - \frac{k^{2}} {2}h_{+}  \;  ,
\end{equation}
\begin{equation}
R_{0102} = R_{3132} = - R_{0132} = - R_{3102} = - \frac{k^{2}} {2} 
h_{\times}  \;  ,
\end{equation}
and all the other independent components are xero.

\subsection{Fermi Normal Coordinates}
Manasse and Misner\cite{MM} constructed Fermi normal coordinates to 
second order in the displacement from a point in free fall.  We 
generalize their approach here by deriving the coordinates to all 
orders in the displacement from a freely falling center of mass, but 
to first order in the gravitational wave amplitude.  

Starting with a point on the CM's world line with proper time $\tau$, 
we run spacelike geodesics out a distance 
$\lambda$.  These geodesics start in the direction $\alpha^{i}$ , i = 
1, 2, 3, so any point P in a neighborhood of the CM can be considered 
to have the coordinates 
\begin{equation}
(x^{0} , x^{i}) = (\tau, \lambda\alpha^{i})  \; .
\end{equation}
Also, at the CM, the metric can be taken to be $\eta_{\mu \nu}$  with 
zero first derivatives.

Let $s^{\alpha}$ denote the tangent to such a space like geodesic, 
and let $n^{\alpha}$ denote a deviation vector.    Then the equation 
of geodesic deviation, Manasse and Misner's Eq. (63), becomes
\begin{equation}
\frac{d^{2}n^{\mu}}{ds^{2}} + 
2\frac{dn^{\sigma}}{ds}\Gamma^{\mu}_{\sigma \alpha}s^{\alpha} + 
n^{\sigma}s^{\alpha}s^{\beta}[R^{\mu}_{\alpha \sigma \beta} + 
\Gamma^{\mu}_{\sigma \alpha , \beta} + \Gamma^{\tau}_{\sigma 
\alpha}\Gamma^{\mu}_{\tau \beta} - \Gamma^{\mu}_{\sigma 
\tau}\Gamma^{\tau}_{\alpha \beta}] = 0   \; .   \label{eq:riemann}
\end{equation}
We take the gravitational waves to be small of first order and set
\begin{equation}
R^{\mu}_{\alpha \sigma \beta} \rightarrow R^{\mu}_{\alpha \sigma 
\beta} e^{ik\cdot x}  \;  ,
\end{equation}
taking the real part at the end of the calculation, and, since the 
Riemann tensor is small of first order, we may take in the exponent 
the metric tensor in $k\cdot x$ to be of zeroth order, 
\begin{equation}
k\cdot x = k^{\gamma}x_{\gamma} = k^{\gamma}\eta_{\gamma 
\delta}x^{\delta}  \;  .
\end{equation}
The $\Gamma^{2}$ terms are of second order and may be neglected.

We will denote the metric and its perturbation in Fermi normal 
coordinates as
\begin{equation}
g_{\mu \nu} = \eta_{\mu \nu} + h_{\mu \nu}  \;  ,
\end{equation}
With n a vector between neighboring space-like geodesics,

$n = \frac{\partial}{\partial \alpha^{i}}$  ,  $n^{\sigma} = \lambda\delta^{\sigma}_{i} ,$  
  $\frac{dn^{\sigma}}{d\lambda} = \delta^{\sigma}_{i}$ 
,  $\frac{d^{2}n^{\sigma}}{d\lambda^{2}} = 0$  ,     

and s the tangent to the geodesic,
\begin{equation}
s^{\sigma} = \partial_{\lambda}x^{\sigma} = 
\partial_{\lambda}(\lambda\alpha^{i}\delta^{\sigma}_{i}) = 
\alpha^{i}\delta^{\sigma}_{i}  \;  ,
\end{equation}
Eq. (\ref{eq:riemann}) becomes
\begin{equation}
\lambda\Gamma^{\mu}_{ij,k}\alpha^{j}\alpha^{k} + 
2\Gamma^{\mu}_{ij}\alpha^{j} + 
\lambda\alpha^{j}\alpha^{k}R^{\mu}_{jik}e^{ikx} = 0  \;  ,
\end{equation}
which is just an ordinary differential equation,
\begin{equation}
\partial_{\lambda}(\lambda^{2}\Gamma^{\mu}_{ij}\alpha^{j}) = - 
\lambda^{2}\alpha^{j}\alpha^{k}R^{\mu}_{jik}e^{i\tau}e^{i\lambda\vec{k}\cdot\vec{x}} 
= 0  \;  ,
\end{equation}
and can be integrated to give the $\Gamma^{\mu}_{ij}$ as 
functions of $\lambda$.  Then, using the first order affine condition,
\begin{equation}
\frac{\partial g_{ij}}{\partial\lambda} = \alpha^{k}g_{ij,k} = 
\alpha^{k}[\Gamma^{\mu}\!_{ik}\eta_{\mu j} + 
\Gamma^{\mu}\!_{jk}\eta_{\mu i}]  \;  ,
\end{equation}
we get the space-space portion of the metric.  A similiar process 
can be applied to the other metric components, and the results for TT 
gauge are:
\begin{equation}
h_{00} = - \frac {k^2}{2}[h_{+}(x^{2} - y^{2}) 
+2h_{\times}xy]q(kz)e^{-ik\tau}  \;  ,
\label{time-time}
\end{equation}
and, defining the vector $h^{i} = h_{0i}$, we get
\begin{equation}
\vec{h} = \frac {k^{2}}{3}\left[h_{+}\left(
\begin{array}{c}
- zx \\ 
zy \\ 
x^{2} - y^{2} 
\end{array}
\right) + 
h_{\times}\left(
\begin{array}{c}
- zy \\ 
- zx \\ 
2xy 
\end{array}
\right)\right]p(kz)e^{-ik\tau}  \;  ,
\end{equation}
also
\begin{equation}
(h_{ij}) = \frac {k^2}{2}\left(\begin{array}{ccc}
- h_{+}z^{2}  & - h_{\times}z^{2}  &  (h_{+}x + h_{\times}y)z   \\
- h_{\times}z^{2}  &  h_{+}z^{2}  &  (h_{\times}x - h_{+}y)z    \\
(h_{+}x + h_{\times}y)z   &  (h_{\times}x - h_{+}y)z  &  - 
(h_{+}(x^{2} - y^{2}) + 2h_{\times}xy)  
\end{array}\right)g(kz)e^{-ik\tau}  \;  ,
\label{space-space}
\end{equation}
where
\begin{equation}
g(w) = - \frac {6}{w^{2}} \left[ - 2\frac {\left(e^{iw} - 
1\right)}{iw} + e^{iw} +1 \right]  \;  ,
\label{gweq}
\end{equation}
\begin{equation}
p(w) = \frac {3}{2}\left[\frac {\left(e^{iw} - 1\right)}{(iw)^{2}} - 
\frac {1}{iw} + \frac {g(w)}{6}\right]  \;  ,
\end{equation}
\begin{equation}
q(w) = 2\left[\frac {\left(e^{iw} - 1\right)}{(iw)^{2}} - \frac 
{1}{iw}\right]  \;  .
\label{qweq}
\end{equation}
These weight functions all $\rightarrow 1$ in the long wavelength 
limit, and then Eqs. (\ref{time-time}) - (\ref{space-space}) become identical to Manasse and Misner's 
results.  In this limit
\begin{equation}
    \vec{\nabla}\cdot\vec{h} = \nabla^{2}h_{00} = 0  \;  .
    \label{longwave}
\end{equation}

\section{Fields in Conductors}
\subsection{Charged Particles in Curved Space}
The Lagrangian for a particle of charge q, mass m, interacting with 
the electromagnetic field in a curved space is:
\begin{equation}
L = - mc\left[ 
-g_{\alpha\beta}\dot{x}^{\alpha}\dot{x}^{\beta}\right]^{\frac{1}{2}} + 
qg_{\alpha\beta}\dot{x}^{\alpha}A^{\beta}  \; .
\end{equation}
Linearising, this is, for non-relativistic motion,
\begin{equation}
L = - mc^{2}\left[1 - \frac{h_{00}}{2} -  
\frac{\vec{h}\cdot\vec{v}}{c} - \frac{v^{2}}{2c^{2}}\right] -q\Phi + 
q\vec{v}\cdot\vec{A}  \; .
\end{equation}
With both $h_{\mu\nu}$ and $\vec{A}$ small of first order, the canonical 
conjugate momentum is
\begin{equation}
\vec{p} = m\vec{v} + mc\vec{h} + q\vec{A}  \;  ,
\end{equation}
and the Hamiltonian is
\begin{equation}
H = \frac{1}{2m}\left[\vec{p} - q\vec{A} -mc\vec{h}\right]^{2} + 
q\Phi - \frac{mc^{2}}{2}h_{00}  \; .
\label{hamiltonian}
\end{equation}
This implies that an electron with $q = -e$ in a resistive medium with 
collision frequency $\tau$ would have the equation of motion
\begin{equation}
m\dot{\vec{v}} + \frac{m\vec{v}}{\tau} = -e\vec{E} - mc\dot{\vec{h}} + 
\frac{mc^{2}}{2}\vec{\nabla} h_{00}  \; .
\end{equation}
The normal electron current is then
\begin{equation}
\vec{J_{n}} = \sigma_{n}\left[\vec{E} + \frac{mc}{e}\dot{\vec{h}} - 
\frac{mc^{2}}{2e}\vec{\nabla} h_{00}\right]  \;  ,
\end{equation}
where for frequency $\omega << \frac{1}{\tau}$ the normal electron conductivity is
\begin{equation}
\sigma_{n} = \frac{e^{2}\tau n_{n}}{m}  \;  ,
\end{equation}
where $n_{n}$ is the normal electron density.

\subsection{Field Solutions}
For both the normal and super-conductors, the field 
equations inside the conductors are of the form
\begin{equation}
\nabla^{2}\phi - \kappa^{2}\phi = f(\vec{x})  \;  ,
\end{equation}
where $\phi$ is a field amplitude, $f(\vec{x})$ is the source proportional to the 
gravitational field, and $\frac{1}{\kappa} = \lambda$ is either the 
skin depth, for normal conductors, or the London 
penetration depth, for superconductors.  This depth can vary from mm 
in normal conductors to some 100 nm in superconductors, and can be 
expected to be much smaller than both the scale $\lambda_{W}$ over which the gravitational wave 
varies significantly and also much smaller than the conductor's size R. 
The general solution for $\phi$ is 
\begin{equation}
\phi(\vec{x}) = - \int d^{3}x'\frac {e^{-\kappa |x - x'|}}{4\pi|x - 
x'|}f(x') + \phi_{H}(x)  \;  ,
\end{equation}
where $\phi_{H}(x)$ is a homogeneous solution.  Under our scale 
assumptions this reduces to 
\begin{equation}
\phi(\vec{x}) = - \frac{f(\vec{x})}{\kappa^{2}} + 
\phi_{H}e^{\kappa\zeta}  + O\left(\frac{\lambda}{\lambda_{W}}\right)  \;  ,
\end{equation}
where $\zeta$ denotes the distance from the conductor surface, (taken 
as positive in the outwards direction) and $\phi_{H}$ is independent 
of $\zeta$ but may depend on the coordinates parallel to the surface.

The amplitude for the homogeneous solution has to be determined by 
fitting the interior solution to an outgoing wave at the boundary.

\subsection{Fields in a Normal Conductor}
We assume that the gravitational field oscillates with frequency 
$\omega$, then
\begin{equation}
\vec{J_{n}} = \sigma_{n}\left[\vec{E} - i\omega \frac{mc}{e}\vec{h} - 
\frac{mc^{2}}{2e}\vec{\nabla} h_{00}\right]  \; .
\end{equation}
With a typical value of $ \sigma_{n}\sim 10^{8}/\Omega m$ and with 
$\omega < 10^{8}Hz$ , we can neglect the displacement current, and 
Ampere's law becomes
\begin{equation}
\vec{\nabla}\times\vec{B} = \mu_{0}\sigma_{n}\left[\vec{E} - i\omega \frac{mc}{e}\vec{h} - 
\frac{mc^{2}}{2e}\vec{\nabla} h_{00}\right]  \; .
\label{amp-norm}
\end{equation}
This implies
\begin{equation}
\vec{\nabla}\cdot\vec{E} = i\omega 
\frac{mc}{e}\vec{\nabla}\cdot\vec{h} + 
\frac{mc^{2}}{2e}\nabla^{2}h_{00}  \; .
\label{gauss-norm}
\end{equation}
Eq. (\ref{amp-norm}) implies 
\begin{equation}
    \nabla^{2}\vec{B} + i\omega\mu_{0}\sigma_{n}\vec{B} = 
    i\omega\mu_{0}\sigma_{n}\frac{mc}{e}\vec{\nabla}\times\vec{h}  \; . 
\end{equation}
For $10^{3}Hz < \omega < 10^{8}Hz$ the skin depth
\[\lambda_{S} = \sqrt{\frac{2}{\omega\mu_{0}\sigma_{n}}}  \; , \]
runs between 5 mm and 10 $\mu m$ and is much smaller than the 
gravitational wavelength.  With the conductor's size on the order of a 
meter, the considerations of the previous section apply, and 
\begin{equation}
    \vec{B} = \frac{mc}{e}\vec{\nabla}\times\vec{h} + 
    \vec{B}_{H}e^{\kappa \zeta}  \;  ,
\end{equation}
where $\kappa = \frac{1 - i}{\lambda_{S}}$. 
The time derivative of Eq. (\ref{amp-norm}), Faraday's law 
and Eq. (\ref{gauss-norm}) lead to 
\begin{equation}
    \vec{E} = i\omega \frac{mc}{e}\vec{h} + 
\frac{mc^{2}}{2e}\vec{\nabla} h_{00} + 
\frac{1}{i\omega\mu_{0}\sigma_{n}}\vec{\nabla}\cdot\vec{E} + 
\vec{E}_{H}e^{\kappa \zeta}  \; .
\label{E-gen-sol}
\end{equation}

When the gravitational wavelength is much larger than the conductor 
size, Eq. (\ref{longwave}) applies, leading to $\vec{\nabla}\cdot\vec{E} = 0$ and
\begin{equation}
    \vec{E} = i\omega \frac{mc}{e}\vec{h} + 
\frac{mc^{2}}{2e}\vec{\nabla} h_{00} + \vec{E}_{H}e^{\kappa \zeta} \; .
\label{E-norm-long}
\end{equation}

\subsection{The Superconductor}

The Ginzburg-Landau free energy for a superconductor with 
supercarriers of mass $m^{\ast} = 2m$ ,  charge $e^{\ast} = -2e$, is\cite{KS},
\begin{equation}
    F = 
    \int_{V}d^{3}x\left[\frac{1}{2m^{\ast}}\left|\left(-i\hbar\vec{\nabla} - 
    e^{\ast}\vec{A}\right)\Psi\right|^{2} + 
    e^{\ast}\Phi\left|\Psi\right|^{2} - \alpha\left|\Psi\right|^{4} + 
    \frac{\beta}{2}\left|\Psi\right|^{4}\right] \; .
    \end{equation}
Comparision with the Hamiltonian of Eq. (\ref{hamiltonian}) suggests 
that it would be reasonable to extend this to an effective Lagrangian,
\begin{equation}
L = \int_{V}d^{3}x\left[i\hbar\Psi^{\ast}\frac{\partial\Psi}{\partial t} - 
\cal{H}_{S}\right] \;  ,
\end{equation}
where
\begin{equation}
\cal{H}\mit_{S} = \frac{1}{2m^{\ast}}\left|\left(-i\hbar\vec{\nabla} - e^{\ast}\vec{A}- m^{\ast}c\vec{h}\right)\Psi\right|^{2} + 
    \left(e^{\ast}\Phi - 
    \frac{m^{\ast}c^{2}}{2}h_{00} \right)\left|\Psi\right|^{2} 
    - \alpha\left|\Psi\right|^{2} + 
    \frac{\beta}{2}\left|\Psi\right|^{4} \; .
\end{equation}
Varying the corresponding action leads to the equation for the order parameter $\Psi$,
\begin{equation}
 i\hbar\frac{\partial\Psi}{\partial t} = 
\left[\frac{1}{2m^{\ast}}\left(-i\hbar 
 \vec{\nabla} - e^{\ast}\vec{A}- m^{\ast}c\vec{h}\right)^{2} +  
 e^{\ast}\Phi - 
 \frac{m^{\ast}c^{2}}{2}h_{00} -\alpha + \beta 
 \left|\Psi\right|^{2}\right]\Psi \;  ,
 \label{psieq}
\end{equation}
and a boundary condition at the superconductor surface,
\begin{equation}
    \hat{n}\cdot\left(-i\hbar\vec{\nabla} - e^{\ast}\vec{A} - 
    m^{\ast}c\vec{h}\right)\Psi = 0 \; .
    \label{fullBC}
    \end{equation}
The supercurrent is
\begin{equation}
    \vec{J}_{S} = - \frac{ie^{\ast}\hbar}{2m^{\ast}}\left(\Psi^{\ast}\vec{\nabla}\Psi - 
    \vec{\nabla}\Psi^{\ast}\Psi\right) - \frac{e^{\ast 2}}{m^{\ast}}\left(\vec{A} + 
    \frac{m^{\ast}c}{e^{\ast}}\vec{h}\right)\left|\Psi\right|^{2}
    \label{supJ}
    \end{equation}
and the boundary condition Eq. (\ref{fullBC}) implies that at the 
surface
\begin{equation}
    \hat{n}\cdot\vec{J}_{S} = 0  \; .
    \label{currentBC}
    \end{equation}
Linearising, 
\begin{equation}
    \Psi = \left|\Psi_{0}\right|\left(1 + s\right)e^{i\theta} \;  ,
\end{equation}
and introducing the gauge invariant phase $\tilde{\theta}$,
\begin{equation}
    \theta = \tilde{\theta} - \frac{e^{\ast}}{\hbar}\int\Phi dt \;  ,
    \end{equation}
Eq. (\ref{psieq}) yields the linearised equations
\begin{equation}
    i\hbar\omega\tilde{\theta} = - 
    \frac{{\hbar}^{2}}{2m^{\ast}}\nabla^{2}s - \frac{m^{\ast}c^{2}}{2}h_{00} + 
    2\alpha s  \; ,
    \label{thetaeq}
    \end{equation}
\begin{equation}
   i\hbar\omega s = \frac{1}{2m^{\ast}}\left[{\hbar}^{2}\nabla^{2}\tilde{\theta} + 
   \hbar
   \vec{\nabla}\cdot\left(\frac{e^{\ast}}{-i\omega}\vec{E} - 
   m^{\ast}c\vec{h}\right)\right] \;  ,
   \label{sequation}
   \end{equation}
and the boundary conditions at the surface,
\begin{equation}
    \hat{n}\cdot\left(\hbar\vec{\nabla}\tilde{\theta} + 
    \frac{e^{\ast}}{-i\omega}\vec{E} - m^{\ast}c\vec{h}\right) = 0  \;  ,
\end{equation}
\begin{equation}
    \hat{n}\cdot\vec{\nabla}s= 0  \; .
    \label{sbc}
    \end{equation}
The London penetration depth is, 
\begin{equation}
    \lambda_{L} = \sqrt{\frac{m^{\ast}}{\mu_{0}(e^{\ast})^{2}
    \left|\Psi_{0}\right|^{2}}}  \;  ,
    \end{equation}
and in our generic superconductor, $ \lambda_{L} \sim 10^{-7}m$ . We also 
use the London frequency, 
\begin{equation}
    \omega_{L} = \frac{1}{\lambda_{L}\sqrt{\mu_{0}\epsilon}} \;  ,
    \end{equation}
which is about $10^{15}Hz$.
Linearized, the current of Eq. (\ref{supJ}) is now
\begin{equation}
\vec{J}_{S} = \frac{1}{\mu_{0}e^{\ast}\lambda_{L}^{2}}\left(\hbar
\vec{\nabla}\tilde{\theta} + \frac{e^{\ast}\vec{E}}{-i\omega} -  
m^{\ast}c\vec{h}\right) \; .
\end{equation}
With no normal electrons present, Gauss' law becomes
\begin{equation}
    \vec{\nabla}\cdot\vec{E} = 
    \frac{2e^{\ast}\left|\Psi_{0}\right|^{2}s}{\epsilon} = 
    \frac{2m^{\ast}\omega_{L}^{2}}{e^{\ast}}s  \; .
    \label{supergauss}
\end{equation}
and with this Eq. (\ref{sequation}) becomes
\begin{equation}
   i\hbar\omega s = \frac{\hbar^{2}}{2m^{\ast}}\nabla^{2}\tilde{\theta} + 
   i\hbar\frac{\omega_{L}^{2}}{\omega}s 
    - \frac{\hbar c}{2}\vec{\nabla}\cdot\vec{h}  \; .
   \label{seq2}
   \end{equation}
In the long wavelength limit, the divergence of $\vec{h}$ can be neglected, 
moreover, $ \frac{\omega_{L}^{2}}{\omega^{2}} > 10^{14} $, for 
$\omega > 10^{3}Hz$ , so Eq. (\ref{seq2}) reduces to 
\begin{equation}
    s = i\xi^{2}_{L}\nabla^{2}\tilde{\theta}  \;  ,
\label{seq3} 
\end{equation}
where
\begin{equation}
    \xi_{L} = \sqrt{\frac{\hbar\omega}{2m^{\ast}\omega_{L}^{2}}}  \; .
    \end{equation}
This varies from $10^{-16}$ to $10^{-13} m$ as $\omega$ goes from 
 $10^{3}$ to $10^{8} Hz$.
Using this in Eq. (\ref{thetaeq}) leads to
\begin{equation}
    i\frac{\hbar\omega}{\alpha}\tilde{\theta} = 
    -i\xi^{2}\xi_{L}^{2}\nabla^{4}\tilde{\theta} - 
    \frac{m^{\ast}c^{2}}{2\alpha}h_{00} + 2i\xi_{L}^{2}\nabla^{2}\tilde{\theta} \;  ,
    \label{theta2}
    \end{equation}
where
\begin{equation}
    \xi = \sqrt{\frac{\hbar^{2}}{2m^{\ast}\alpha}}  \;  ,
    \end{equation}
is the Ginzburg-Landau coherence length, $\sim 10^{-7}m$ in our 
generic superconductor.

Deep inside the superconductor, a depth much greater than the 
characteristic lengths $\xi$, $\xi_{L}$, we will have
\begin{equation}
    \tilde{\theta} = i\frac{m^{\ast}c^{2}}{2\hbar\omega}h_{00} \;  ,
    \end{equation}
and $s \rightarrow 0$, since $\nabla^{2}h_{00} = 0$ in the long 
wavelength limit.
Near the surface $\tilde{\theta}$ will be of the form
\begin{equation}
    \tilde{\theta} =  i\frac{m^{\ast}c^{2}}{2\hbar\omega}h_{00} + 
    \tilde{\theta}_{0}e^{iq\zeta}  \;  ,
    \end{equation}
where q satisfies 
\begin{equation}
    i\frac{\hbar\omega}{\alpha} = -i\xi^{2}\xi_{L}^{2}q^{4} + 
    2i\xi_{L}^{2}q^{2}  \; .
    \end{equation}
The roots corresponding to attenuation as one goes into the conductor 
are
\begin{equation}
    q_{\pm} = \frac{1}{\xi}\left[1 \pm \sqrt{\frac{\xi^{2}\hbar\omega}{\xi^{2}_{L}\alpha} - 
    1}\right]^{\frac{1}{2}}  \;  ,
    \end{equation}
for our generic superconductor, 
\begin{equation}
    q_{\pm} = \sqrt{\pm i}q_{0}  \;  ,
    \end{equation}
where
\begin{equation}
    q_{0} = \left[\frac{2m^{\ast}c^{2}}{\alpha}\right]^{\frac{1}{4}}
    \frac{1}{\sqrt{\xi\lambda_{L}}}  \; .
\end{equation}
Now
\begin{equation}
    \tilde{\theta} = i\frac{m^{\ast}c^{2}}{2\hbar\omega}h_{00} + 
    \tilde{\theta}_{+}e^{q_{+}\zeta} + 
    \tilde{\theta}_{-}e^{q_{-}\zeta} \;  ,
    \end{equation}
and, by Eq. (\ref{seq3}), 
\begin{equation}
    s = - \xi^{2}_{L}q_{0}^{2}\left[\tilde{\theta}_{+}e^{q_{+}\zeta} - 
    \tilde{\theta}_{-}e^{q_{-}\zeta}\right]  \; .
    \end{equation}
The boundary condition, Eq. (\ref{sbc}) implies that
\begin{equation}
     \tilde{\theta}_{-} = i \tilde{\theta}_{+}  \;  ,
     \end{equation}
so we can write
\begin{equation}
    \tilde{\theta} = i\frac{m^{\ast}c^{2}}{2\hbar\omega}h_{00} + 
    \tilde{\theta}_{H}\left(e^{q_{+}\zeta} + ie^{q_{-}\zeta}\right)  \;  ,
    \label{tilde-sol}
    \end{equation}
and
\begin{equation}
    s = - \xi^{2}_{L}q_{0}^{2}\tilde{\theta}_{H}\left[e^{q_{+}\zeta} - 
    ie^{q_{-}\zeta}\right]  \; .
    \end{equation}

\subsection{Fields in the Superconductor}

Neglecting the displacement current, Ampere's law is now
\begin{equation}
    \vec{\nabla}\times\vec{B} = 
    \frac{1}{\lambda_{L}^{2}e^{\ast}}\left(\hbar\vec{\nabla}\tilde{\theta} + 
    \frac{e^{\ast}E}{-i\omega} - m^{\ast}c\vec{h}\right)  \;  ,
    \label{amplaw}
    \end{equation}
taking the curl, this becomes
\begin{equation}
    \nabla^{2}\vec{B} - \frac{1}{\lambda_{L}^{2}}\vec{B} =  
    \frac{m^{\ast}c}{e^{\ast}\lambda_{L}^{2}}\vec{\nabla}\times\vec{h} \;  ,
    \end{equation}
which has the solution,
\begin{equation}
    \vec{B} = - \frac{m^{\ast}c}{e^{\ast}}\vec{\nabla}\times\vec{h} + 
    \vec{B}_{H}e^{\kappa\zeta}  \;  ,
    \label{bsol}
    \end{equation}
where $\kappa = \frac{1}{\lambda_{L}}$.
Now Eqs. (\ref{amplaw}) and (\ref{bsol}) give for the E field,
\begin{equation}
    \vec{E} = 
    i\frac{\omega}{e^{\ast}}\left(\hbar\vec{\nabla}\tilde{\theta} -  
    m^{\ast}c\vec{h}\right) 
     -i\omega\lambda_{L}^{2}\vec{\nabla}\times\left(\vec{B}_{H}e^{\kappa\zeta}\right) + 
     \frac{i\omega\lambda_{L}^{2}m^{\ast}c}{e^{\ast}}
     \vec{\nabla}\times\left(\vec{\nabla}\times\vec{h}\right)  \; . 
\label{Esol}
 \end{equation}
In the long wavelength limit the last term in this equation will be 
zero.  In practice, it is convenient to first write E in the 
Eq. (\ref{Esol}) form, obtain B from Faraday's equation, and then 
obtain the surface terms by fitting to an outward propagating 
radiation field.

It should be noted that with the use of Eq. (\ref{tilde-sol}), in the 
long wavelength limit the interior part of Eq. (\ref{Esol}) becomes 
identical to that of Eq. (\ref{E-norm-long}) with the substitutions 
$e \rightarrow - e^{\ast}$ and $m \rightarrow m^{\ast}$.

\section{Example: The Sphere}
\subsection{Vector Spherical Harmonics}

Let $\vec{L} = -i\vec{x}\times\vec{\nabla}$, then the vector $\vec{h}$ 
can be written as
\begin{equation}
    \vec{h} = \frac{ik^{2}r^{2}}{6}\vec{L}C(\theta,\phi) \;  ,
    \label{hCeq}
    \end{equation}
where the function
\begin{equation}
    C(\theta,\phi) = sin^{2}(\theta)\left[h_{+}sin(2\phi) - 
    h_{\times}cos(2\phi)\right]  \;  ,
    \end{equation}
is a superposition of $\ell = 2$ spherical harmonics.
Defining a similiar function
\begin{equation}
    D(\theta,\phi) = sin^{2}(\theta)\left[h_{+}cos(2\phi) + 
    h_{\times}sin(2\phi)\right]  \;  ,
    \end{equation}
we can write
\begin{equation}
    \vec{\nabla}h_{00} = 
    \vec{\nabla}\times\frac{ik^{2}r^{2}}{6}\vec{L}D  \; .
    \label{h00Deq}
    \end{equation}

A useful vector identity is, for an arbitrary function $f(r)$, and an 
$\ell = 2$ spherical harmonic $Y_{2,m}(\theta,\phi)$,
\begin{equation}
    \vec{\nabla}\times\vec{L}f(r)Y_{2,m)(\theta,\phi} = 
    \left[\left(f' + \frac{f}{r}\right)\hat{n}\times\vec{L} + 
    \frac{6i\hat{n}}{r}f\right]Y_{2,m}(\theta,\phi)  \; .
    \end{equation}

\subsection{The Normal Conductor}

We now write the E field of Eq. (\ref{E-norm-long}) as
\begin{equation}
    \vec{E} = - \frac{\omega mck^{2}}{6e}r^{2}\vec{L}C +     
    \frac{imc^{2}k^{2}}{12e}\vec{\nabla}\times\left(r^{2}\vec{L}D\right) +
    E_{0C}e^{\kappa(r - R}\vec{L}C - 
    \frac{i\omega}{\kappa^{2}}B_{0D}\vec{\nabla}\times
    \left(e^{\kappa(r - R}\vec{L}D\right) \;  ,
    \end{equation}
where we have taken the surface terms as having the same angular 
dependence as the volume gravitational source terms.

This gives, using Faraday's law and neglecting terms of order 
$(\delta/R)$,
\begin{equation}
    \vec{B} = 
    \frac{imck^{2}}{6e}\vec{\nabla}\times\left(r^{2}\vec{L}C\right)
    + \frac{E_{0C}}{i\omega}\vec{\nabla}\times\left(e^{\kappa(r - 
    R}\vec{L}C\right) + B_{0D}e^{\kappa(r - R}\vec{L}D  \; .
    \end{equation}
At the surface, the D wave produces a transverse B field, and the C 
wave produces a transverse E field.  We therefore take for the 
exterior fields an outward radiating D wave TM field, and a C wave TE 
field\cite{jackson},
\begin{equation}
    \vec{E} = E_{1}h^{1}_{2}(kr) + 
    \frac{B_{1}c^{2}}{-i\omega}\vec{\nabla}
    \times\left(h^{1}_{2}(kr)\vec{L}D\right)  \;  ,
    \label{outerE}
    \end{equation}
\begin{equation}
    \vec{B} = \frac{E_{1}}{i\omega}\vec{\nabla}
    \times\left(h^{1}_{2}(kr)\vec{L}C\right)
    + B_{1}h^{1}_{2}(kr)\vec{L}D  \; .
    \label{outerB}
    \end{equation}

\subsection{At the Surface}
When $r = R$, the inner E field becomes
\begin{equation}
    \vec{E} = - \frac{\omega mck^{2}}{6e}R^{2}\vec{L}C +
    \frac{imc^{2}k^{2}}{4e}R\left(\hat{n}\times\vec{L} + 
    2i\hat{n}\right)D +  E_{0C}\vec{L}C -
     \frac{i\omega}{\kappa^{2}}B_{0D}\left(\kappa\hat{n}
     \times\vec{L} + \frac{6i\hat{n}}{R}\right)D  \;  ,
     \end{equation}
and the outer E field is
\begin{equation}
    \vec{E} = E_{1}h^{1}_{2}(kR) + \frac{B_{1}c^{2}}{-i\omega}\left[
    F\hat{n}\times\vec{L} + \frac{6i\hat{n}}{R}h^{1}_{2}(kR)\right]D
    \label{surfaceOutE}  \;  ,
    \end{equation}
where
\begin{equation}
    F = kh^{1\prime}_{2}(kR) + \frac{h^{1}_{2}(kR)}{R}  \; .
    \end{equation}
In the following we will abbreviate $h^{1}_{2}(kR) = g$.
Continuity of tangential E implies that
\begin{equation}
    - \frac{\omega mck^{2}}{6e}R^{2} + E_{0C} = E_{1}g  \;  ,
    \label{E-LC}
    \end{equation}
\begin{equation}
    \frac{imc^{2}k^{2}}{4e}R - \frac{i\omega}{\kappa}B_{0D} = 
    \frac{B_{1}c^{2}}{-i\omega}F  \; .
    \label{E-nxLD}
    \end{equation}
    
The inner B field at the surface becomes
\begin{equation}
    \vec{B} = \frac{imck^{2}}{2e}R\left(\hat{n}\times\vec{L} + 
    2i\hat{n}\right)C + \frac{E_{0C}}{i\omega}
    \left(\kappa\hat{n}\times\vec{L} + \frac{6i\hat{n}}{R}\right)C
    + B_{0D}\vec{L}D \;  ,
    \end{equation}
and the outer field is
\begin{equation}
    \vec{B} = \frac{E_{1}}{i\omega}\left[
    F\hat{n}\times\vec{L} + \frac{6i\hat{n}}{R}g\right]C
    + B_{1}g\vec{L}D  \; .
    \label{surfaceOutB}
    \end{equation}
Continuity of tangential B leads to 
\begin{equation}
    \frac{imck^{2}}{2e}R + \frac{E_{0C}}{i\omega} = 
    \frac{E_{1}}{i\omega}F  \;  ,
    \end{equation}
\begin{equation}
    B_{0D} = B_{1}g  \; .
    \end{equation}
Continuity of normal B leads to an equation identical to Eq. (\ref{E-LC}).
The amplitude coefficients are therefore:
\begin{equation}
    E_{0C} = - \frac{\omega mck^{2}R^{2}}{6e\kappa}\left[F - 
    \frac{3g}{R}\right]\left[g - \frac{F}{\kappa}\right]^{-1} \;  ,
    \label{genampE0C} 
    \end{equation}
\begin{equation}
    B_{0D} =  \frac{\kappa mc^{2}k^{2}R}{4e\omega}\left[1 + \frac{\kappa 
    F}{k^{2}g}\right]^{-1} \;  ,
    \label{genampB0D}
    \end{equation}
\begin{equation}
    E_{1} = - \frac{\omega mck^{2}R}{6e}\left[R - 
    \frac{3}{\kappa}\right]\left[g - \frac{F}{\kappa}\right]^{-1} \;  ,
    \end{equation}
\begin{equation}
    B_{1} =  \frac{\kappa mc^{2}k^{2}R}{4\omega eg}\left[1 + \frac{\kappa 
    F}{k^{2}g}\right]^{-1} \; .
    \end{equation}
We consider here the case when the gravitational wavelength is much 
larger than both the sphere's radius R and the London penetration 
depth $\lambda_{L}$. Then $kR << 1$ and $\frac{\kappa}{k} >> 1$ and 
with
\begin{equation}
    g \rightarrow - \frac{3}{(kR)^{3}} \;  ,
    \end{equation}
\begin{equation}
    F \rightarrow \frac{6}{R(kR)^{3}} \;  ,
    \end{equation}
then
\begin{equation}
    E_{0C} = \frac{5\omega mck^{2}R}{6e\kappa} \;  ,
    \end{equation}
\begin{equation}
    B_{0D} = - \frac{mc^{2}k^{4}R^{2}}{8e\omega} \;  ,
    \end{equation}
\begin{equation}
    E_{1} = \frac{\omega mck^{5}R^{5}}{18e} \;  ,
    \end{equation}
\begin{equation}
    B_{1} =  \frac{mck^{6}R^{5}}{24e} \; .
    \end{equation}

\subsection{The Radiation Field}
At large r, 
\begin{equation}
    h^{1}_{2}(kr) \rightarrow i\frac{e^{ikr}}{kr} \;  ,
    \end{equation}
and Eq. (\ref{outerE}) becomes
\begin{equation}
    \vec{E} \rightarrow iE_{1}\frac{e^{ikr}}{kr}\vec{L}C - 
    iB_{1}\frac{e^{ikr}}{kr}\hat{n}\times\vec{L}D  \; .
    \end{equation}
The radiation power distribution is
\begin{equation}
    \frac{dP}{d\omega} = \frac{|r\vec{E}|^{2}}{2c\mu_{0}}  \; .
    \end{equation}
With an unpolarised beam, 
\begin{equation} 
    <h_{+}^{2}> = <h_{\times}^{2}> = \frac{h^{2}}{2}  \;  ,
\end{equation}
\begin{equation}
<h_{+}h_{\times}> = 0  \;  ,
\end{equation}
this becomes
\begin{equation}
    \frac{dP}{d\omega} = \frac{25c}{2592\mu_{0}}
    \left(\frac{mc }{e}\right)^{2}(kR)^{10}h^{2}sin^{2}(\theta)\left[cos^{2}
    (\theta) - \frac{48}{25}cos(\theta) + 1\right]  \;  ,
    \end{equation}
and the total radiated power is
\begin{equation}
    P = \frac{5\pi c}{162\mu_{0}}\left(\frac{mc} 
    {e}\right)^{2}(kR)^{10}h^{2}  \; .
    \end{equation}
The incident power flux in the gravitational wave is\cite{weinberg},
\begin{equation}
    J_{0} = \frac{k^{2}c^{5}h^{2}}{16\pi G}  \;  ,
    \label{J0eq}
    \end{equation}
so the efficiency of gravitational to electromagnetic wave conversion is
\begin{equation}
    \eta = \frac{P}{J_{0}\pi R^{2}} 
         = \frac{40\pi 
         G}{81\mu_{0}}\left(\frac{m}{ce}\right)^{2}(kR)^{8} 
	 \sim 3\times 10^{-44}(kR)^{8}  \;  ,
    \end{equation}
which for spheres of meter radius is a rather small number.

\subsection{The Superconducting Sphere}

The driving forces for the fields are proportional to the $\ell = 2$ 
spherical harmonics C and D.  We therefore set
\begin{equation}
    \tilde{\theta} = - i\frac{m^{\ast}c^{2}k^{2}r^{2}D}{4\hbar\omega} + 
    \left(\tilde{\theta}_{C}C + 
    \tilde{\theta}_{D}D\right)\left(e^{q_{+}\zeta} + 
    ie^{q_{-}\zeta}\right)  \; .
    \end{equation}
Then
\begin{equation}
    \vec{\nabla}\tilde{\theta} = - i\frac{m^{\ast}c^{2}k^{2}
    \vec{\nabla}\times r^{2}\vec{L}D}{12\hbar\omega} +
    \sqrt{i}q_{0}\hat{n}\left(\tilde{\theta}_{C}C + 
    \tilde{\theta}_{D}D\right)\left(e^{q_{+}\zeta} + 
    e^{q_{-}\zeta}\right) + O\left(\frac{1}{qR}\right)  \; .
    \end{equation}
The source of the electric field has both an $\vec{L}C$ and a 
$\vec{\nabla}\times\vec{L}D$ term.  We therefore include such terms in 
the harmonic solution, and take for the total electric field
\begin{eqnarray}
    \vec{E} = - i\frac{m^{\ast}c^{2}k^{2}
    \vec{\nabla}\times r^{2}\vec{L}D}{12e^{\ast}} + 
    i\sqrt{i}\omega\frac{\hbar}{e^{\ast}}q_{0}\hat{n}\left(\tilde{\theta}_{C}C + 
    \tilde{\theta}_{D}D\right)\left(e^{q_{+}\zeta} +  
    e^{q_{-}\zeta}\right)  + \nonumber \\
    \frac{\omega m^{\ast}c}{6e^{\ast}}k^{2}r^{2}\vec{L}C +
    E_{0C}e^{\kappa\zeta}\vec{L}C - i\omega\lambda^{2}_{L}B_{0D}
    \vec{\nabla}\times\left(e^{\kappa\zeta}\vec{L}D\right)  \;  ,
    \end{eqnarray}
from this we get, using Faraday's law,
\begin{equation}
    \vec{B} = -i\frac{m^{\ast}c}{6e^{\ast}}k^{2}\vec{\nabla}\times
    \left(r^{2}\vec{L}C\right) + \frac{E_{0C}}{i\omega}
    \vec{\nabla}\times\left(e^{\kappa\zeta}\vec{L}C\right) +
    B_{0D}e^{\kappa\zeta}\vec{L}D  \; .
    \end{equation}
We have now an $\vec{L}C$ TE mode plus a TM $\vec{L}D$ mode.  We 
therefore take the same combination for the exterior fields, as given 
in Eqs. (\ref{outerE}) and (\ref{outerB}).

At r = R, the inner E field becomes
\begin{eqnarray}
    \vec{E} = - i\frac{m^{\ast}c^{2}k^{2}R}{4e^{\ast}}\left(\hat{n}
    \times\vec{L} + 2i\hat{n}\right)D + 
    2i\sqrt{i}\omega\frac{\hbar}{e^{\ast}}q_{0}\hat{n}\left(\tilde{\theta}_{C}C + 
    \tilde{\theta}_{D}D\right) +  \nonumber \\
    \frac{\omega m^{\ast}c}{6e^{\ast}}k^{2}R^{2}\vec{L}C +
     E_{0C}\vec{L}C - i\omega\lambda_{L}B_{0D}\hat{n}\times\vec{L}D
     + O\left(\frac{\lambda_{L}}{R}\right) \;  ,
     \end{eqnarray}
and the inner B field becomes
\begin{equation}
    \vec{B} = -i\frac{m^{\ast}c}{2e^{\ast}}k^{2}R\left(\hat{n}
    \times\vec{L} + 2i\hat{n}\right)C + \frac{E_{0C}}{i\omega}
    \left(\kappa\hat{n}\times\vec{L} + \frac{6i\hat{n}}{R}\right)C +
    B_{0D}\vec{L}D  \; .
    \end{equation}
At the surface, $\vec{B}$ and the parallel component of $\vec{E}$ 
must be continuous.  Moreover, since the normal component of the 
supercurrent vanishes at the surface, there can be no surface charge, 
and as a result Gauss' law requires that
\begin{equation}
    \hat{n}\cdot\vec{E}_{in} = \frac{\epsilon_{0}}{\epsilon}
    \hat{n}\cdot\vec{E}_{out}
    \label{gaussE}  \; .
    \end{equation}
Parallel E conductivity gives for the coefficients of the independent 
surface functions:
\begin{equation}
    \frac{\omega m^{\ast}c}{6e^{\ast}}k^{2}R^{2} + E_{0C} = E_{1}g \;  ,
    \label{LCeq}
    \end{equation}
\begin{equation}
    - i\frac{m^{\ast}c^{2}k^{2}R}{4e^{\ast}} - i\omega\lambda_{L}B_{0D}
    = - \frac{c^{2}}{i\omega}B_{1}F \; .
    \end{equation}
Eq. (\ref{gaussE}) for the normal component of E gives
\begin{equation}
    \tilde{\theta}_{C} = 0  \;  ,
    \end{equation}
\begin{equation}
    \frac{m^{\ast}c^{2}k^{2}R}{2e^{\ast}} +
    2i\sqrt{i}\omega\frac{\hbar}{e^{\ast}}q_{0}\tilde{\theta}_{D}
    = - \frac{6\epsilon_{0}c^{2}}{\omega\epsilon}B_{1}\frac{g}{R}  \; .
    \end{equation}
Parallel B continuity implies:
\begin{equation}
    B_{0D} = B_{1}g \;  ,
    \end{equation}
\begin{equation}
    \frac{m^{\ast}ck^{2}R}{2e^{\ast}} + \frac{\kappa E_{0C}}{\omega}
    = \frac{E_{1}}{\omega}F  \; .
    \end{equation}
Continuity of the normal component of B leads to an equation identical 
to Eq. (\ref{LCeq}).

The coefficients for the E and B fields are the same as in 
Eqs. (\ref{genampB0D}) to (\ref{genampE0C}) with the substitutions $m 
\rightarrow m^{\ast}$ and $-e \rightarrow e^{\ast}$.  The new equation 
here is
\begin{equation}
     \tilde{\theta}_{D} = - \frac{\sqrt{i}m^{\ast}c^{2}k^{2}R}{4\hbar\omega
     q_{0}}\left[\frac{3\kappa}{\epsilon_{r}k^{2}R}
     \left(1 + \frac{\kappa F}{k^{2}g}\right)^{-1} - 1\right] \;  ,
     \end{equation}
where $\epsilon_{r}$ is the superconductor's dielectric constant.

The E and B field amplitudes are the same as for the normal conductor, with the appropriate mass, 
charge and penetration depth, but the phase amplitude becomes in the 
long wave limit
\begin{equation}
    \tilde{\theta}_{D} \rightarrow \frac{\sqrt{i}m^{\ast}c^{2}k^{2}R}
    {4\hbar\omega q_{0}}\left(\frac{3}{2\epsilon_{r}} + 1\right) \; .
    \end{equation} 
The external fields are essentially the same as for the normal 
conductor.  The efficiency of GW to EMW conversion is just as 
astronomically small.

The phase $\tilde{\theta}$ is of some interest. The bulk value is
\begin{equation}
    |\tilde{\theta}| \sim  
    \frac{m^{\ast}c^{2}}{2\hbar\omega}(kR)^{2}h  \; .
    \end{equation}
Then with $h \sim 10^{-24}$, $R \sim 1m$ and waves in the megacycle 
range, $|\tilde{\theta}| \sim 10^{-14}$.
At the surface it changes by a negligible amount,
\begin{equation}
    |\tilde{\theta}_{D}D| \sim \frac{m^{\ast}c(kR)^{3}}{\hbar q_{0}}h
    \sim 10^{-19}  \; .
    \end{equation}
The phase is therefore essentially equal everywhere to its bulk value 
of about $10^{-14}$, a small but not astronomically small quantity, 
which could perhaps be measurable.

\section{High Frequency Gravitational Waves}

There is a possibility that high frequency gravitational waves may 
exist as relics of the Big Bang\cite{hfgw}.  We show below that 
for waves in the gigacycle range, the conversion efficiency is still 
extremely low, but the phase amplitude can be somewhat larger.  

The amplitude functions of Eqs. (\ref{gweq}) - (\ref{qweq}) fall off with large kz.  Thus 
to maximise output at high frequency we require a target of small 
extent in the z-direction.  We consider therefore a disk 
perpendicular to the x-axis, extending from z = -a to +a, and of 
radius R.  We take
\begin{equation}
    k\lambda_{L} << ka << 1 << kR  \; .
    \end{equation}
Although non-zero, it turns out that $\vec{\nabla}\cdot\vec{h}$ and 
$\nabla^{2}h_{00}$ are still negligible compared to the other terms.
Let
\begin{equation}
    \vec{u} = x\hat{x} - y\hat{y} \;  ,
    \end{equation}
\begin{equation}
    \vec{v} = y\hat{x} + x\hat{y}  \; .
    \end{equation}
We have, neglecting the surface variation in $\tilde{\theta}$, and 
considering only the transverse E and B components, inside the 
superconductor near the $z = +a$ side,
\begin{equation}
    \vec{E}_{T} =  \frac{m^{\ast}c^{2}k^{2}}{2e^{\ast}}\left[h_{+}\vec{u}
    + h_{\times}\vec{v}\right] + \left[E_{0u}\vec{u} + 
    E_{0v}\vec{v}\right]e^{\kappa\left(a - z\right)}  \;  ,
    \end{equation}
\begin{equation}
    \vec{B}_{T} = - \frac{m^{\ast}c^{2}k^{2}}{e^{\ast}}\left[- 
    h_{+}\vec{v} + h_{\times}\vec{u}\right] - \frac{\kappa}{i\omega}
    \left[E_{0u}\vec{vu} - E_{0v}\vec{u}\right]e^{\kappa\left(a - 
    z\right)}  \; .
    \end{equation}
(Here we have also neglected the part of E coming from $\vec{h}$ as it 
is negligible here in comparision to the contribution from 
$\vec{\nabla}\tilde{\theta}$.)  We approximate the external fields on 
the positive z side as 
\begin{equation}
    \vec{E} = \left[E_{1u}\vec{u} + E_{1v}\vec{v}\right]e^{ik\left(z - 
    a\right)} \;  ,
    \end{equation}
\begin{equation}
    \vec{B} = \frac{1}{c}\left[E_{1u}\vec{v} - 
    E_{1v}\vec{u}\right]e^{ik\left(z - a\right)}  \; .
    \end{equation}
The boundary conditions lead to
\begin{equation}
    E_{1u} =  \frac{m^{\ast}c^{2}k^{2}}{2e^{\ast}}h_{+} + O(k\lambda_{L}) \;  ,
    \end{equation}
\begin{equation}
    E_{1v} =  \frac{m^{\ast}c^{2}k^{2}}{2e^{\ast}}h_{\times} + 
    O(k\lambda_{L}) \;  ,
    \end{equation}
and $E_{0u}$ and $E_{0v}$ are similiar but smaller by a factor of $k\lambda_{L}$.
For an unpolarised wave, the power radiated from the positive side is
\begin{equation}
    P = \frac{\pi c}{24\mu_{0}}\left(\frac{m^{\ast}c}{e^{\ast}}\right)^{2}
    (kR)^{4}h^{2} \;  ,
    \end{equation}
and the efficiency becomes
\begin{equation}
    \eta = \frac{\pi}{2}\left(\frac{m^{\ast}c}{e^{\ast}}
    \right)^{2}G(kR)^{2} \; .
    \end{equation}
Now $\eta \sim 3 \times 10^{-44}(kR)^{2}$ and is still extremely small, even 
for $R \sim 10 m$ with frequencies in the 100 Gigacycle range, 
where $kR \sim 10^{4}$. However, the phase $\tilde{\theta}$ is now
\begin{equation}
    \tilde{\theta} = -i\frac{m^{\ast}c^{2}}{4\hbar \omega}k^{2}
    \left(h_{+}\left(x^{2} - y^{2}\right) + 2h_{\times}xy\right) \;  ,
    \end{equation}
and
\begin{equation}
    |\tilde{\theta}| \sim 10^{12}kR^{2}h   \;  .
    \end{equation}
This would be $\sim 10^{-6}$ for $h \sim 10^{-24}$ . However, Eq. 
(\ref{J0eq}) gives for a wave with such an amplitude a flux of $\sim 
10^{9} W/m^{2}$ !  A more reasonable estimate of the cosmic background 
at this frequency would be $h \sim 10^{-32}$ \cite{hfgw}, and then 
$|\tilde{\theta}| \sim 10^{-14}$ , the same as for the low frequency 
wave.   
\section{Conclusion}

We find that the conversion of gravitational wave energy to electromagnetic 
energy by both normal and super-conductors is a remarkably 
inefficient process. However in the superconducting case, the 
gravitational wave produces a perturbation in the order parameter phase 
that may perhaps be detectable.


\begin{thebibliography}{}
\bibitem{chaio} R. Y. Chiao, arXiv:gr-qc/0204012, arXiv:gr-qc/0208024.
\bibitem{z} S. A. Zhou, \textit{Electrodynamic Theory of 
Superconductors}, (Peter Perigrinus Ltd., 1991).
\bibitem{weinberg} S. Weinberg, \textit{Gravitation And Cosmology}, 
(John Wiley, 1972).
\bibitem{MM} F. K. Manasse and C. W. Misner, J. Math. Phys. 
\textbf{4}, 735 (1963).
\bibitem{KS} J. B. Ketterson and S. N. Song, 
\textit{Superconductivity}, (Cambridge University Press, 1999).
\bibitem{jackson} J. D. Jackson, \textit{Classical Electrodynamics}, 
(John Wiley, 1975).
\bibitem{hfgw} A. Buonanno, arXiv:gr-qc/0303085.
\end{thebibliography}
\end{document}